\documentclass[
reprint,
onecolumn,
superscriptaddress,
showkeys,
amsmath, 
amssymb,  
longbibliography, 
]{revtex4-1}

\usepackage{graphicx}
\usepackage{dcolumn}
\usepackage{bm}

\begin{document}

\title{Enhanced Superconducting Transition Temperature due to Tetragonal Domains in Two-Dimensionally Doped SrTiO$_{3}$}

\author{Hilary Noad}
\affiliation{Stanford Institute for Materials and Energy Sciences, SLAC National Accelerator Laboratory, 2575 Sand Hill Road, Menlo Park, CA 94025, USA}
\affiliation{Department of Applied Physics, Stanford University, Stanford, CA 94305, USA}
\author{Eric M. Spanton}
\affiliation{Stanford Institute for Materials and Energy Sciences, SLAC National Accelerator Laboratory, 2575 Sand Hill Road, Menlo Park, CA 94025, USA}
\affiliation{Department of Physics, Stanford University, Stanford, CA 94305, USA}
\author{Katja C. Nowack}
\affiliation{Department of Applied Physics, Stanford University, Stanford, CA 94305, USA}
\author{Hisashi Inoue}
\affiliation{Department of Applied Physics, Stanford University, Stanford, CA 94305, USA}
\author{Minu Kim}
\affiliation{Stanford Institute for Materials and Energy Sciences, SLAC National Accelerator Laboratory, 2575 Sand Hill Road, Menlo Park, CA 94025, USA}
\author{Tyler A. Merz}
\affiliation{Stanford Institute for Materials and Energy Sciences, SLAC National Accelerator Laboratory, 2575 Sand Hill Road, Menlo Park, CA 94025, USA}
\affiliation{Department of Applied Physics, Stanford University, Stanford, CA 94305, USA}
\author{Christopher Bell}
\affiliation{Stanford Institute for Materials and Energy Sciences, SLAC National Accelerator Laboratory, 2575 Sand Hill Road, Menlo Park, CA 94025, USA}
\affiliation{HH Wills Physics Laboratory, University of Bristol, Tyndall Avenue, Bristol, BS8 1TL, UK}
\author{Yasuyuki Hikita}
\affiliation{Stanford Institute for Materials and Energy Sciences, SLAC National Accelerator Laboratory, 2575 Sand Hill Road, Menlo Park, CA 94025, USA}
\author{Ruqing Xu}
\affiliation{Advanced Photon Source, Argonne National Laboratory, Argonne, IL 60439, USA}
\author{Wenjun Liu}
\affiliation{Advanced Photon Source, Argonne National Laboratory, Argonne, IL 60439, USA}
\author{Arturas Vailionis}
\affiliation{Stanford Institute for Materials and Energy Sciences, SLAC National Accelerator Laboratory, 2575 Sand Hill Road, Menlo Park, CA 94025, USA}
\author{Harold Y. Hwang}
\affiliation{Stanford Institute for Materials and Energy Sciences, SLAC National Accelerator Laboratory, 2575 Sand Hill Road, Menlo Park, CA 94025, USA}
\affiliation{Department of Applied Physics, Stanford University, Stanford, CA 94305, USA}
\author{Kathryn A. Moler}
\affiliation{Stanford Institute for Materials and Energy Sciences, SLAC National Accelerator Laboratory, 2575 Sand Hill Road, Menlo Park, CA 94025, USA}
\affiliation{Department of Applied Physics, Stanford University, Stanford, CA 94305, USA}
\affiliation{Department of Physics, Stanford University, Stanford, CA 94305, USA}
\email{kmoler@stanford.edu}

\begin{abstract}
Strontium titanate is a low-temperature, non-Bardeen-Cooper-Schrieffer superconductor that superconducts to carrier concentrations lower than in any other system and exhibits avoided ferroelectricity at low temperatures. Neither the mechanism of superconductivity in strontium titanate nor the importance of the structure and dielectric properties for the superconductivity are well understood. We studied the effects of twin structure on superconductivity in a 5.5-nm-thick layer of niobium-doped SrTiO$_{3}$ embedded in undoped SrTiO$_{3}$. We used a scanning superconducting quantum interference device susceptometer to image the local diamagnetic response of the sample as a function of temperature. We observed regions that exhibited a superconducting transition temperature $T_{c}$ $\agt$ 10\% higher than the temperature at which the sample was fully superconducting. The pattern of these regions varied spatially in a manner characteristic of structural twin domains. Our results emphasize that the anisotropic dielectric properties of SrTiO$_{3}$ are important for its superconductivity, and need to be considered in any theory of the mechanism of the superconductivity.
\end{abstract}

\maketitle

\section{\label{sec:introduction}Introduction}
Superconductivity in electron-doped SrTiO$_{3}$ (STO) most likely arises from electron-phonon coupling \cite{KooncePR67,AppelPR69,ZinamonPhilosophMag70,NgaiPRL74,TakadaJPSJ80,BaratoffPhysB81,KliminPRB12,EdgePRL15,GorkovARXIV15,RosensteinARXIV16}, yet it cannot be described by conventional Bardeen-Cooper-Schrieffer (BCS) theory \cite{BardeenPR57} because the Fermi temperature in STO is lower than the Debye temperature, opposite to the requirement of BCS. Certain features of the superconductivity are reminiscent of unconventional, high-temperature superconductors: there is a dome-like dependence of transition temperature ($T_{c}$) on doping \cite{SchooleyPRL65,KooncePR67,LinPRX13,LinPRL14} that is superficially similar to the domes found in the cuprates and iron pnictides \cite{BrounNatPhys08,ShibauchiARCMP14}. Further, STO superconductivity occurs close to a ferroelectric quantum critical point \cite{MullerPRB79,RowleyNatPhys14,EdgePRL15}; quantum criticality is important for superconductivity in the cuprates \cite{BrounNatPhys08} and the iron pnictides \cite{ShibauchiARCMP14}. Superconductivity in STO is a puzzle in and of itself, and is also important in the context of understanding superconductivity in thin-film and interfacial systems that are grown on STO. Furthering our understanding of STO superconductivity may shed light on the role of STO in the reported pseudogap behavior of LaAlO$_{3}$ (LAO)/STO heterostructures \cite{RichterNat13,ChengNat15}. It may also illuminate the contribution of STO phonons and the importance of the dielectric properties of STO to monolayer FeSe on STO \cite{HeNatMater13,TanNatMater13,LeeNat14}. 

A cubic-to-tetragonal structural phase transition occurs in STO at 105 K: small rotations of TiO$_{6}$ octahedra cause the unit cell to double in height and the in-plane axes to rotate by 45$^{\circ}$ and lengthen by a factor of $\sqrt{2}$ \cite{UnokiJPSJ67}. We will refer to the tetragonal unit cell using the pseudocubic convention ($a$ = $a_{tet}/\sqrt{2}$ and is oriented parallel to cubic $\langle$100$\rangle$; $c$ = $c_{tet}/2$). The tetragonal crystal phase allows three orientations of crystallographic twin domains to form. The twins are distinguished by whether the tetragonal $c$ axis points along the former cubic [100], [010], or [001] axis. The twin structure strongly influences local normal-state electronic properties  \cite{KaliskyNatMater13,HonigNatMater13} and weakly modulates the superfluid density at temperatures well below $T_{c}$ in LAO/STO heterostructures \cite{KaliskyNatMater13}. By studying the effects of the perturbation due to twin structure on superconducting $\delta$-doped STO, we hope to expand our understanding of the origin of superconductivity in STO.
 
Here, we studied the effects of twin structure on superconductivity in $\delta$-doped STO \cite{KozukaNat09,KozukaAPL10a,KimPRL11,KimPRB12}. Using a scanning superconducting quantum interference device (SQUID) susceptometer, we observed a local enhancement in $T_{c}$ that was set by the tetragonal twin structure of the material. We believe that the underlying mechanism of this enhancement in $T_{c}$ is related to modulations of the dielectric environment that are driven by the orientation of the lattice relative to the superconducting plane and by generic alterations of the dielectric environment near twin boundaries.

\section{\label{sec:experimental}Experimental setup}
We used a SQUID susceptometer in a dilution refrigerator with a base temperature below 50 mK (during scanning) \cite{BjornssonRevSciInstrum01,HuberRevSciInstrum08} to study superconductivity in $\delta$-doped STO \cite{KozukaNat09,KozukaAPL10a,KimPRL11,KimPRB12}. The SQUID susceptometer consists of a gradiometric SQUID whose pickup loops ($\approx$ 3 $\mu$m in diameter) are arranged concentrically with single-turn field coils ($\approx$ 20 $\mu$m in diameter). We measured the response of the sample using the primary pickup loop-field coil pair, while the counterwound rear pickup loop canceled the response of the SQUID to the applied field \cite{HuberRevSciInstrum08}. 

We rastered the SQUID over the sample in a plane parallel to the $\delta$-doped layer in the STO and spatially mapped the diamagnetic response of the superconductor to the magnetic field that we applied with the field coil. In the sensor geometry that we used for our measurements, the diamagnetic response of the two-dimensional superconductor is directly proportional to the superfluid density (Appendix~\ref{sec:ns}) \cite{KirtleyPRB12,BertPRB12}. We observed the critical temperature locally by determining when the diamagnetism disappeared (when the measured susceptibility matched a background measurement). 

We studied two samples of $\delta$-doped STO as well as a single-crystal sample of bulk Nb-doped STO (dopant concentration $N_D$ = 1 at.\%) that was obtained from Shinkosha, Inc. The $\delta$-doped samples were fabricated using pulsed laser deposition, with the growth conditions described elsewhere \cite{KozukaAPL10}. The structures consist of a thin layer (thickness $d$) doped with Nb that is embedded between undoped STO cap and buffer layers \cite{KozukaAPL10a,KozukaAPL10} and exhibits two-dimensional superconductivity \cite{KozukaNat09}. The electrons are confined around the doped layer by the Coulomb potential from the dopant ions.

This investigation focused on the results from a thinner and higher density $\delta$-doped sample ($d$ = 5.5 nm, $N_D$ = 1 at.\% Nb), but we also measured a thicker and lower density $\delta$-doped sample ($d$ = 36.9 nm, $N_D$ = 0.2 at.\% Nb). We measured the 1 at.\% Nb $\delta$-doped sample in two separate cooldowns, the second one occurring after having warmed the sample to room temperature, removed it to a desiccator from its sample holder, and stored it for several months.

The total area of the $d$ = 5.5 nm, $N_{D}$ = 1 at.\% Nb sample was approximately 7.7 mm$^{2}$. In the first cooldown, we imaged $\approx$ 30\% of the total area at a temperature close to but generally below the upper $T_{c}$ \cite{SupplementalInfo}. Imaging in this temperature range allowed us to identify areas of interest for studying the temperature dependence of the susceptibility while efficiently exploring the sample over millimeter length scales. The total area over which we imaged the temperature dependence of the susceptibility was ${\approx} 2.8{\times}10^{5}$ $\mu$m$^{2}$ in the first cooldown and ${\approx} 1.6{\times}10^{5}$ $\mu$m$^{2}$ in the second, or approximately 4\% and 2\% of the total sample area, respectively. 

We performed differential aperture X-ray microdiffraction \cite{microdiff1,microdiff2,microdiff3} experiments at beamline 34-ID-E at the Advanced Photon Source, Argonne National Laboratory \cite{Xraypaper} on the 1 at.\% Nb $\delta$-doped STO. This beamline is equipped with a liquid nitrogen-cooled stage that we used to cool the STO below its structural transition temperature of 105 K \cite{UnokiJPSJ67}. We collected Laue diffraction patterns while rastering the sample under the X-ray beam, then indexed each pattern to a distorted room temperature cubic unit cell for STO \cite{microdiff1,microdiff2,microdiff3} in order to determine the orientation of the local crystal structure (extended discussion in Appendix~\ref{sec:microdiff}). 

\section{\label{sec:results}Results}

\begin{figure*}
\includegraphics{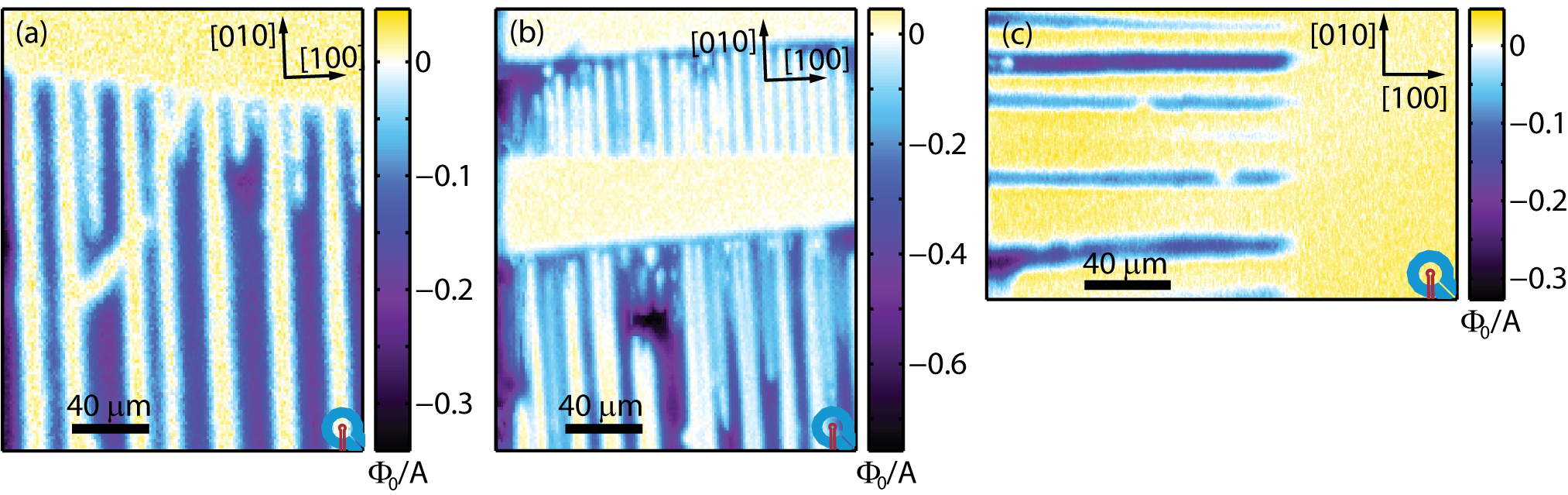}
\caption{\label{fig:1}Color. In $\delta$-doped STO, long, narrow regions oriented along cubic [100] and [010] are superconducting at temperatures at which their surroundings are no longer superconducting. (a-c), Maps of magnetic susceptibility in different areas of a single $\delta$-doped sample reveal patterns of superconductivity along [010] (a,b) and [100] (c). Negative susceptibility (in purple and blue) indicates that a region is superconducting. Yellow regions are in the normal state but become superconducting at lower temperatures. Scans were taken at (a) 330 mK, (b) 300 mK, and (c) 320 mK. The schematics of the SQUID pickup loop (red) and field coil (blue) are to scale and are oriented as they were during data acquisition.}
\end{figure*}

To determine the spatial dependence of $T_{c}$, we mapped the susceptibility as a function of temperature near $T_{c}$ in several regions of the 1 at.\% Nb $\delta$-doped STO sample [Fig.~\ref{fig:1}(a)-(c)]. Some parts of the scanned areas were diamagnetic, indicating that their $T_{c}$ was higher than the scan temperature. In contrast, surrounding parts had zero or very weakly positive (paramagnetic) susceptibility, indicating that they were not superconducting and that their $T_{c}$ was lower than or equal to the scan temperature. We observed similar regions of enhanced $T_{c}$ in the 0.2 at.\% Nb $\delta$-doped STO sample (Appendix~\ref{sec:similar features}). 

\begin{figure}
\includegraphics{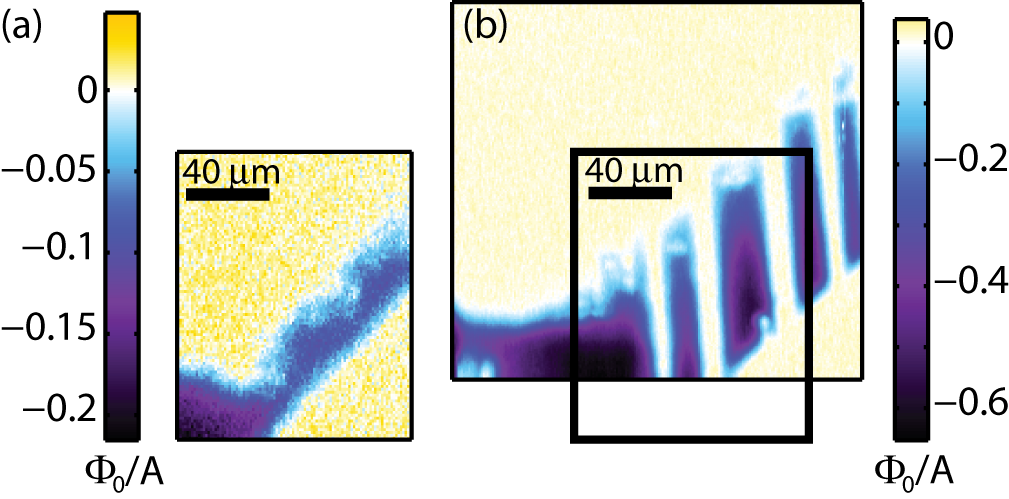}
\caption{\label{fig:2}Color. Superconducting features change spatially after thermal cycling, indicating that they are most likely due to the cubic-to-tetragonal transition in STO. The same region of a single sample was imaged at 340 mK before (a) and after (b) cycling to room temperature. Distinct features are apparent in (b) in a region that lacked sharp features in (a). The box in (b) indicates the approximate location of the image taken in (a). During sample growth, a clip masked the paramagnetic region in the lower right of the image.}
\end{figure}

\begin{figure*}
\includegraphics{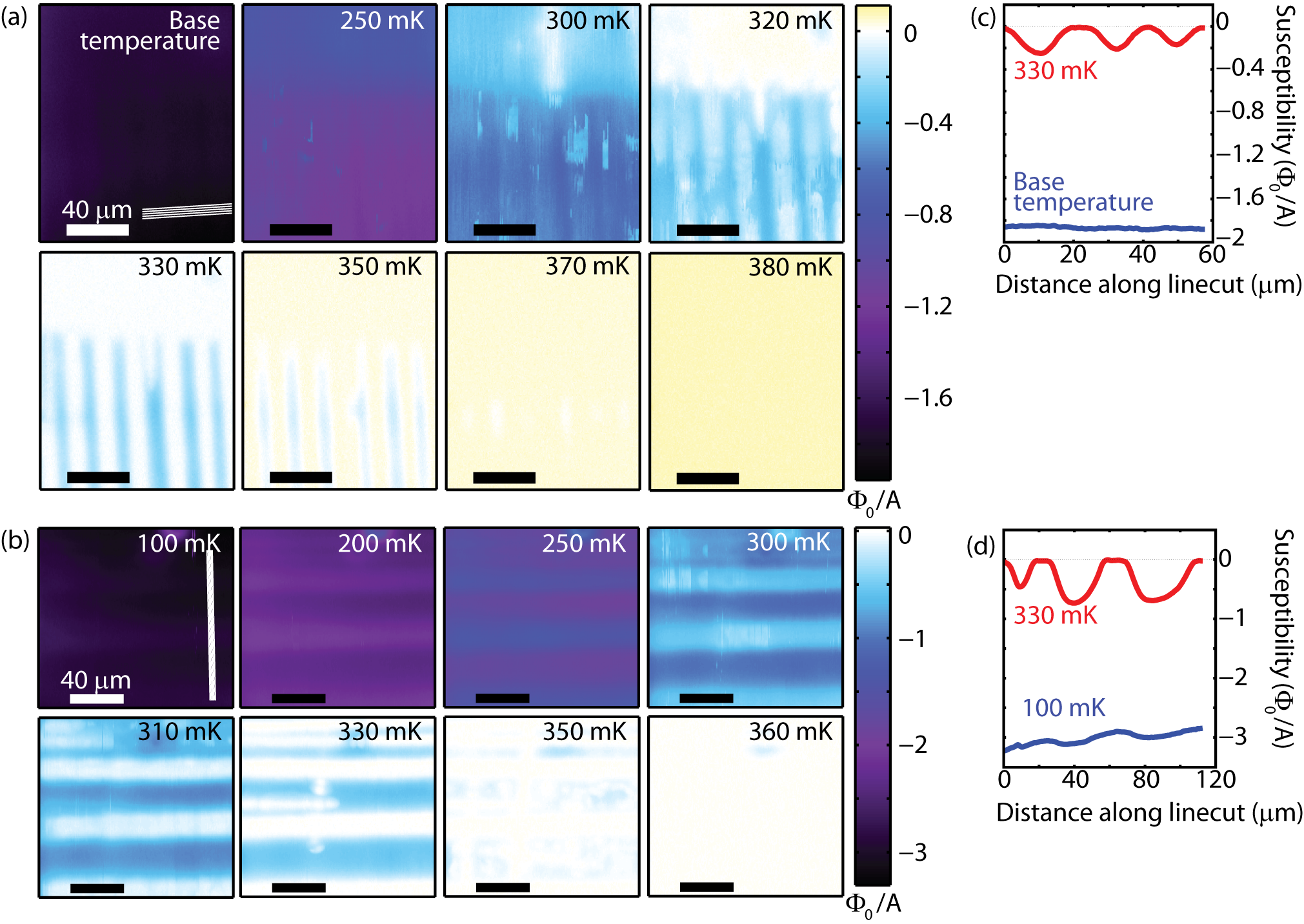}
\caption{\label{fig:3}Color. The spatial variation in susceptibility near $T_{c}$ is much weaker at lower temperatures. (a) Repeated scans at a single location. (b) Repeated scans at a second location. White lines in (a) and (b) indicate the positions of linecuts displayed in (c) and (d), respectively. (c,d) Averaged linecuts taken at 330 mK and at base temperature (c) or at 330 mK and 100 mK (d) demonstrate that the relative amplitude of modulation of the superconducting response is larger at temperatures near $T_{c}$ than at the lowest temperatures.}
\end{figure*}

\begin{figure}
\includegraphics{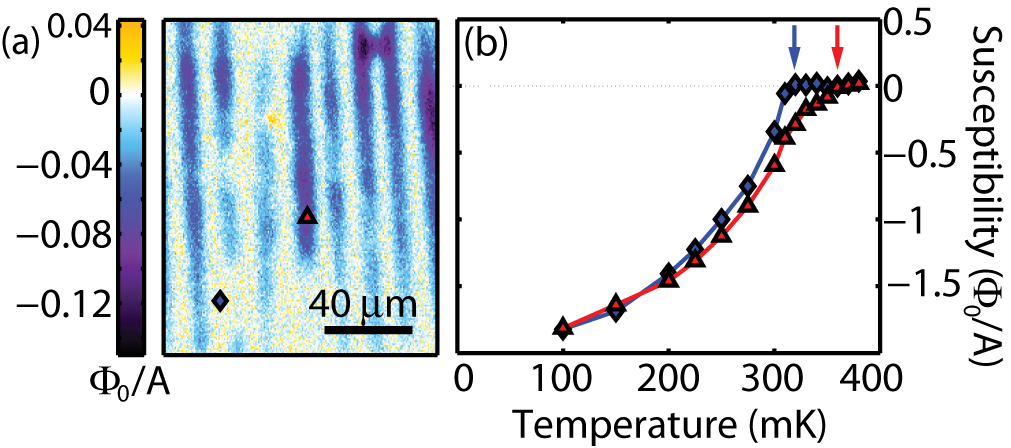}
\caption{\label{fig:4}Color. The temperature dependence of the susceptibility indicates a $\approx$ 10\% spatial variation in $T_{c}$. (a) Representative image at 350 mK from a series of images taken in the same area at 15 temperatures. The scan area presented here is different from the ones shown in Figures 1-3. (b) Susceptibility as a function of temperature at two locations marked by a blue diamond and a red triangle in (a). The transition temperatures are $\approx$ 320 mK and $\approx$ 360 mK.}
\end{figure}

The patterns we observed in susceptibility images [e.g. Fig.~\ref{fig:1}, Fig.~\ref{fig:2}(b), Fig.~\ref{fig:3}(a)-(b), Fig.~\ref{fig:4}(a)] are consistent with enhanced $T_{c}$ on twin boundaries or on certain tetragonal domains of $\delta$-doped STO. We detected regions of enhanced $T_{c}$ aligned along axes that corresponded to the high-temperature cubic $\langle$100$\rangle$ directions, as determined via comparison to SQUID images that included an oriented edge of the sample. The spacing, splitting, and comb-like structures resemble patterns in images of tetragonal domains in STO taken with a polarized light microscope at higher temperatures \cite{SawaguchiJPSJ63,LytleJApplPhys64,KaliskyNatMater13,HonigNatMater13}.  

Some well-defined features, such as the diagonal mark in the middle-left of Fig.~\ref{fig:1}(a), were evident along other directions. $T_{c}$ was not enhanced on those features, and we believe that these features may be due to damage to the sample. In addition, along the edge of an area that was masked by a clip during growth, we observed diffuse regions that had a higher $T_{c}$ overall [masked area is in the lower right of Fig.~\ref{fig:2}(a)]. The general enhancement of $T_{c}$ in the diffuse regions could be due to differences in growth conditions, strain relaxation, or other unknown effects along the edge of the masked region. 

The configuration of SrTiO$_{3}$ tetragonal domains was previously shown to change on thermal cycling \cite{KaliskyNatMater13}. To test whether the regions of enhanced $T_{c}$ also behaved in this manner, we measured the same 1 at.\% Nb $\delta$-doped sample after warming it above the structural transition at 105 K. We used a region that had been masked by a clip during pulsed laser deposition growth to identify specific positions before and after warming. Before warming, we detected diffuse regions of enhanced $T_{c}$ close to the clipped region, with no sharply defined rectangular features elsewhere [Fig.~\ref{fig:2}(a)]. We observed similar diffuse regions of enhanced $T_{c}$ along the edges of all areas of the clipped region that we imaged in the first cooldown. After warming, we obtained a qualitatively different susceptibility image at the same location and temperature [Fig.~\ref{fig:2}(b)]: the image obtained after warming contained sharp rectangular regions similar to those depicted in Fig.~\ref{fig:1}. The observation that well-defined, $\langle$100$\rangle$-oriented features appeared in this area after warming above the structural transition [Fig.~\ref{fig:2}(b)] strongly suggests that such features originate in the tetragonal domain structure of STO. 

In the first cooldown of the $N_{D}$ = 1 at.\% Nb $\delta$-doped sample, we imaged $\approx$ 30\% of the total sample area at an intermediate temperature and observed $\langle$100$\rangle$-oriented features of enhanced $T_{c}$ in $\approx$ 50\% of the area surveyed \cite{SupplementalInfo}. Although features of enhanced $T_{c}$ were not rare in that particular cooldown, factors such as the cooling rate, unintentional strain from sample mounting, or the geometry of the sample~\cite{MullerSolidStateCommun70} could alter the shape, number, and orientation of tetragonal domains that spontaneously form upon cooling through 105 K~\cite{UnokiJPSJ67}.

To confirm the shape and orientation of structural domains in the 1 at.\% Nb $\delta$-doped STO, we used differential aperture X-ray microdiffraction to obtain real space maps of tilts in the $c$ axis above and below the structural phase transition (Appendix~\ref{sec:microdiff}). The spatial resolution of the X-ray microdiffraction measurements ($\approx$ 5 $\mu$m) was similar to the limit on the scanning SQUID measurements set by the diameter of the pickup loop ($\approx$ 3 $\mu$m). Below the transition, long, narrow features were evident (Fig.~\ref{fig:microdiff}), with the orientations expected for structural domains. Their widths, on the order of tens of microns, were comparable to the features detected via scanning SQUID in $\delta$-doped STO (e.g. Fig.~\ref{fig:1},~\ref{fig:2},~\ref{fig:3},~\ref{fig:4} and Appendix~\ref{sec:widths}). 

To investigate the relationship between local $T_{c}$ and low-temperature superfluid density, we measured the temperature dependence of the susceptibility in a series of images (Fig.~\ref{fig:3}). In the region shown in Fig.~\ref{fig:3}(a), parts of the scanned area were no longer superconducting at $T_{c,\textnormal{low}} \approx$  320 mK, while oriented regions remained superconducting until $T_{c,\textnormal{high}} \approx$ 370 mK. Using the relation ${\Delta}T_{c}/T_{c}$ = 100\%$\times$($T_{c,\textnormal{high}}-T_{c,\textnormal{low}}$)/$T_{c,\textnormal{low}}$, we determined that ${\Delta}T_{c}/T_{c} \approx$ 16\% for the area in Fig.~\ref{fig:3}(a), and ${\Delta}T_{c}/T_{c} \approx$ 9\% for the area in Fig.~\ref{fig:3}(b). Similar estimates for the regions displayed in Fig.~\ref{fig:1}, Fig.~\ref{fig:2}(b), and Fig.~\ref{fig:4} are presented in Appendix~\ref{sec:Tcs}, Table~\ref{tab:Tcs}. Linecuts taken near $T_{c}$ and at temperatures well below $T_{c}$ [Fig.~\ref{fig:3}(c) and (d)] demonstrate that although there was large spatial variation in the susceptibility near $T_{c}$, the variation fell to $\approx$ 5\% of the average signal at 100 mK [Fig.~\ref{fig:3}(b)]. Further, there was little to no modulation of the susceptibility ($\alt$ 2\%) at the lowest temperatures measured [Fig.~\ref{fig:3}(a)]. 

The temperature dependence of the superfluid density can give insight into the nature of the superconducting order parameter. The superconductivity of the $\delta$-doped samples that we measured was in a limit where the superfluid density is directly proportional to the susceptibility that we measured with our SQUID \cite{KirtleyPRB12,BertPRB12}. Therefore, to track the temperature dependence of the superfluid density in our samples, we extracted susceptibility versus temperature curves at two different locations from images of the same area at different temperatures (Fig.~\ref{fig:4}). At the lowest temperatures, the susceptibility begins to flatten [Fig.~\ref{fig:4}(b)], consistent with $s$-wave behavior and inconsistent with a nodal gap, which would yield superfluid density related to temperature in a linear to quadratic fashion depending on scattering \cite{AnnettPRB91}. Note that the direct proportionality between susceptibility and superfluid density \cite{KirtleyPRB12,BertPRB12} is only strictly valid for an infinite sheet geometry, so we cannot draw detailed conclusions from the high temperature functional form of these data. The apparent shoulder in the higher-$T_{c}$ region in Fig.~\ref{fig:4}(b) (red triangles) is most likely due to the geometry of the diamagnetic source changing from a quasi-uniform, infinite plane at low temperatures to a series of separated, narrow strips near $T_{c}$. 

We observed considerable variation in the widths of the regions of enhanced $T_{c}$ (Fig.~\ref{fig:widths}; extended discussion in Appendix~\ref{sec:widths}). Notably, features ranged from a width that was apparently resolution limited [e.g. Fig.~\ref{fig:1}(b)] to a full width at half maximum that was wider than the diameter of the field coil on our SQUID ($\approx$ 20 $\mu$m), [e.g. Fig.~\ref{fig:3}(c)]. The lower limit on the spatial resolution of our SQUID is set by the diameter of the pickup loop ($\approx$ 3 $\mu$m). This variation of widths is consistent with the sizes of domains observed via polarized light microscopy \cite{SawaguchiJPSJ63,LytleJApplPhys64,KaliskyNatMater13,HonigNatMater13}. In contrast, domain boundaries are predicted to have widths on the order of a few unit cells \cite{CaoPRB90}. Features caused by boundaries would have widths limited by the superconducting coherence length, $\approx$ 100 nm~\cite{KimPRB12}, well below our spatial resolution. Thus, many of the features detected here are likely not produced by domain boundaries in the STO sample, but rather are suggestive of domains with higher $T_{c}$.

The orientations of the regions of enhanced $T_{c}$ also help us to distinguish between features occurring at domain boundaries or within certain domains. The intersection of domain boundaries with the (001) superconducting plane can be oriented along [100], [010], or at $\sim$ 45$^{\circ}$  to [100] (Fig.~\ref{fig:location1}). Boundaries that are at $\sim$ 45$^{\circ}$ to [100] only occur between domains that both have the $c$ axis in-plane, whereas the [100] and [010] boundaries occur between a domain with $c$ in-plane and a domain with $c$ out-of-plane. If domains with $c$ axis out-of-plane have a different $T_{c}$ from ones with $c$ in-plane, we would expect to see rectangles of enhanced $T_{c}$ whose borders were oriented along [100] and [010] but not at $\sim$45$^{\circ}$ to [100] (Fig.~\ref{fig:location1}), and this matches what we observe.

We overwhelmingly observed regions of enhanced $T_{c}$ that were oriented along former cubic [100] and [010] axes. This strongly suggests that the enhancement in $T_{c}$ that we observe is within individual domains, instead of at their boundaries. It is possible that twin boundaries could produce a similar signature: sample geometry \cite{MullerSolidStateCommun70} or unintentional strain could favor twin boundary orientations ([100] or [010]) that would not be distinguished from narrow in-domain features.  We detected a single region of enhanced $T_{c}$ at approximately -45$^{\circ}$ from [100] (Fig.~\ref{fig:location2}). The -45$^{\circ}$ feature suggests that twin boundaries may lead to an enhancement in $T_{c}$ in some circumstances. While only boundaries can yield $\sim$ 45$^{\circ}$ features, the range of widths in the [100]- and [010]-oriented features suggests that many of them originate within certain domains.

The two-dimensionality of the $\delta$-doped material may mean that subbands, which are irrelevant in the three-dimensional material, are important for determining $T_{c}$ or other aspects of the superconductivity. We investigated the importance of subband occupation by measuring a $\delta$-doped STO sample that contained 36.9 nm of 0.2 at.\% Nb (Fig.~\ref{fig:similar features}) and comparing it to the $\delta$-doped STO that had 5.5 nm of 1 at.\% Nb (e.g. Fig.~\ref{fig:1}). The overall temperature scale for superconductivity in the 0.2 at.\% Nb $\delta$-doped sample was lower than that in the 1 at.\% Nb $\delta$-doped sample, consistent with transport measurements of $T_{c}$ (Fig.~\ref{fig:Tcs} inset). We detected features of enhanced $T_{c}$ in the 0.2 at.\% Nb $\delta$-doped sample (Fig.~\ref{fig:similar features}) that were qualitatively similar to those in the 1 at.\% Nb $\delta$-doped sample (e.g. Fig.~\ref{fig:1}). Thus, the local variation of $T_{c}$ does not require a specific occupation or configuration of subbands in order to occur. 

\section{\label{sec:discussion}Discussion}
Various parameters are tuned by the crystal domain structure of the STO, including the direction of the elongated $c$ axis, the local strain \cite{ICSD,Xraypaper}, and the dielectric constant \cite{SakudoPRL71}. By comparing the relative variation of these parameters to the observed variation in $T_{c}$, we sought to identify the most important parameters. Here, ${\Delta}T_{c}/T_{c}$ was on the order of $10^{-1}$, where $T_{c}$ is the temperature at which the whole scan area is superconducting. Given the widths of the features observed (Appendix~\ref{sec:widths}), here we will mainly focus on how physical parameters are tuned within the structural domains themselves, as opposed to on boundaries. 

The change in the lattice constant along the lengthened pseudocubic $c$ axis, ${\lvert}c-a{\rvert}/a$, is on the order of $10^{-3}$ \cite{ICSD}. This relative change is much smaller than the relative change in $T_{c}$ detected here. We would expect strain due to the tetragonal mismatch, and the associated change in $T_{c}$, to be largest at domain boundaries, inconsistent with the majority of our observations. Each domain may have built-in strain due to the neighboring domain configuration or other factors, but we do not expect this strain to greatly exceed the lattice constant change (${\lvert}c-a{\rvert}/a$) or to be uniform over many-micron length scales, as would be required to produce broad features [e.g. Fig.~\ref{fig:3}(b)]. 

In undoped, single-domain STO in the tetragonal phase, the value of the static dielectric constant $\varepsilon$ is enormous at 4 K, on the order of $10^{4}$, and depends on the orientation along which it is measured \cite{SakudoPRL71}. The value along the $a$ axis, ${\varepsilon}_{a} \approx$ 25 000, is over twice as large as the value along the $c$ axis, $\varepsilon_{c} \approx$ 11 000 \cite{SakudoPRL71}. For $\delta$-doped STO, the anisotropy in the dielectric constant implies that the local dielectric constant perpendicular to the two-dimensional superconducting plane depends on the direction of the $c$ axis within the structural domain. Because the change in dielectric constant is large while the change in pseudocubic axes is small, we suspect that the change in $T_{c}$ that we observed may be driven primarily by the dielectric properties of the crystal, either directly or through the associated phonon modes. 

This hypothesis accounts for our observations that enhanced $T_{c}$ occurred within certain domains [e.g. Fig.~\ref{fig:1}(a), ~\ref{fig:3}(b)], but it also accommodates the scenario in which $T_{c}$ is enhanced at domain walls (e.g. Fig.~\ref{fig:location2}). In our two-dimensional system, the local dielectric constant perpendicular to the plane varies from domain to domain, a situation that we propose leads to enhanced $T_{c}$ within certain domains. At the same time, the recent suggestion that domain walls in STO are polar \cite{ScottPRL12}, together with previous observations of twin-modified current flow and surface potential in LAO/STO heterostructures \cite{KaliskyNatMater13,HonigNatMater13}, imply that twin boundaries modify their local dielectric environment. It seems plausible that this alteration could lead to variations in $T_{c}$, which would be at domain boundaries rather than within certain domains. 

Variations or modifications of the dielectric constant within the superconducting plane could alter $T_{c}$ through screening of the Coulomb repulsion. For example, electrons in a domain where the $c$ axis lies in-plane (“a-c domains”) experience a dielectric constant that is the average of ${\varepsilon}_{a}$ and ${\varepsilon}_{c}$, whereas electrons in a domain where the $c$ axis points out-of-plane (“a-a domains”) experience primarily ${\varepsilon}_{a}$. Since ${\varepsilon}_{a}$ is larger than the average of ${\varepsilon}_{a}$ and ${\varepsilon}_{c}$, the Coulomb repulsion between electrons in a-a domains should be more strongly screened than in a-c domains. With stronger screening could come stronger pairing and higher $T_{c}$ in a-a domains.

Variations in screening of the Coulomb interaction may also have implications for the confinement of electrons in the $\delta$-doped layer. In $\delta$-doped STO, the Coulomb potential set up by the ionized dopant cores confines the mobile electrons to a narrow, electronically two-dimensional layer \cite{KozukaNat09,KozukaAPL10a}. Local variations in the dielectric constant, both within the doped layer and in its vicinity, should alter the spatial extent of the electrons in the direction perpendicular to the dopant layer. Tuning of the fraction of electrons that dwelled outside the doped layer could tune scattering or two-dimensional electron density in a pattern set by twin structure.

In a BCS $s$-wave superconductor, non-magnetic scattering would change the low-temperature superfluid density but not $T_{c}$ \cite{AndersonJPCS59}, implying that local changes in conductivity or carrier concentration that were due to disorder would not affect $T_{c}$. However, the independence of $T_{c}$ and disorder depends on a theorem that is not valid for superconductors having a non-retarded pairing interaction (Fermi energy smaller than the phonon cutoff frequency) \cite{FayPRB95}. STO is in or close to the non-retarded limit \cite{LinPRX13}; thus, it is possible that variations in scattering, carrier density, or defects could tune $T_{c}$. We note that in bulk-doped STO, $N_{D}$ = 1 at.\% Nb corresponds essentially to the peak of the dome in $T_{c}$ \cite{SchooleyPRL65,KooncePR67,LinPRX13,LinPRL14}. Neglecting disorder effects, this suggests that in our $\delta$-doped sample with $N_{D}$ = 1 at.\% Nb, changes in carrier density in either direction would cause $T_{c}$ to decrease, not increase.

Structurally driven anisotropy in the Fermi surface could potentially alter $T_{c}$. First-principles calculations of the Fermi surface in \emph{bulk} electron-doped STO showed that the tetragonal anisotropy produces a considerable distortion in the Fermi surface, compressing it along the $c$ direction by as much as 35\%~\cite{TaoARXIV16}. In our $\delta$-doped material, the Fermi surface is dramatically altered from the bulk by confinement in the vertical direction~\cite{KimPRL11}, yet despite the change in dimensionality, the overall scale of Tc in our $N_{D}$ = 1 at.\% Nb $\delta$-doped material is within the range of peak $T_{c}$s observed for material doped in the bulk \cite{SchooleyPRL65,KooncePR67,LinPRX13,LinPRL14}.  It is likely that structural anisotropy within the conducting plane, i.e. the orientation of the $c$ axis relative to the plane, causes additional anisotropy in the Fermi surface. However, the relative insensitivity of $T_{c}$ to the large overall change in Fermi surface suggests that smaller, structurally driven changes in the Fermi surface are not the dominant source of our observed variation in $T_{c}$. 

Our results are relevant to understanding both bulk electron-doped STO and two-dimensional electron systems in STO-based heterostructures. $\delta$-doped STO is representative of heterostructures in the reduced dimensionality of its electron system and in that it has been grown via a method similar to that used to grow heterostructures \cite{OhtomoNat04}. At the same time, the $\delta$-doped STO system simply consists of STO that has been doped with Nb, albeit in an unusual geometry that allows us to access effects related to the structural and dielectric ansiotropy that would otherwise be difficult to observe. Effects occurring \emph{within} domains would be much less evident in bulk STO because there would be no special in- or out-of-plane direction that would distinguish certain domains from others, apart from near the surface of the crystal. Domain boundaries could produce some effect on the superconductivity in bulk STO, but the effect would likely be subtle compared to the strong diamagnetic screening of the bulk material. 

Lin and coworkers  \cite{LinPRB15} recently suggested that there may be enhanced $T_{c}$ at twin boundaries in bulk-doped STO due to the observation that, over a wide range of dopings, the transition to zero resistance occurs at a temperature where the bulk electrons are still normal. We believe that the regions of enhanced $T_{c}$ that we observed in $\delta$-doped STO are primarily located within certain domains, rather than at boundaries. In the bulk STO that we investigated here (Appendix~\ref{sec:3d}), we did not observe sharply defined diamagnetic features surrounded by weak paramagnetism near $T_{c}$ as we observed in the $\delta$-doped STO [e.g. Fig.~\ref{fig:1}, Fig.~\ref{fig:2}(b), Fig.~\ref{fig:3}(a)-(b), Fig.~\ref{fig:4}(a)]. However, we did observe faint rectangular features of higher diamagnetic response surrounded by comparatively weaker diamagnetism near, but fully below, $T_{c}$ of the scanned area [Fig.~\ref{fig:3d}(a)]. These features may be near-surface features occurring within certain domains, or they may be due to twin boundaries.

Enhanced $T_{c}$ on domain boundaries in other systems has been inferred from bulk measurements of niobium and tin, both of which are described by standard BCS theory \cite{KhlyustikovAdvPhys87}. Proposed mechanisms for the enhancement in $T_{c}$ included softening of the phonon spectrum or enhanced electron-phonon coupling (due to atoms being further apart and Coulomb repulsion diminished) at the boundaries \cite{KhlyustikovAdvPhys87}. Although STO superconductivity differs from that in tin and niobium in the details, it is possible that similar mechanisms for enhancing $T_{c}$ could be at play. Twin-boundary-driven enhancements in superconductivity are not limited to conventional electron-phonon superconductors. For example, enhanced superfluid density was observed at twin boundaries in underdoped Ba(Fe,Co)$_{2}$As$_{2}$ \cite{KaliskyPRB10}; however, enhanced $T_{c}$ was not. The mechanism for the enhancement in superfluid density was not known at the time of the previous report. 

Published theories of superconductivity in STO \cite{KooncePR67,AppelPR69,ZinamonPhilosophMag70,NgaiPRL74,TakadaJPSJ80,BaratoffPhysB81,KliminPRB12,EdgePRL15,GorkovARXIV15} or STO-based heterostructures \cite{KliminPRB14,LeeNat14,RosensteinARXIV16} that make reference to microscopic mechanisms all consider an electron-phonon pairing mechanism but differ in their treatment of the electron-phonon interaction. Some consider either soft transverse optical ferroelectric phonons \cite{TakadaJPSJ80,EdgePRL15} or high-energy polar longitudinal optical modes \cite{BaratoffPhysB81,LeeNat14,GorkovARXIV15,RosensteinARXIV16} to be important, while others consider non-ferroelectric optical modes \cite{KooncePR67,AppelPR69,NgaiPRL74}, acoustic modes \cite{ZinamonPhilosophMag70}, or a combination of contributions from acoustic and optical modes \cite{KliminPRB12,KliminPRB14}. Plasmon-mediated electron pairing was additionally considered in the low-density regime \cite{TakadaJPSJ80}. Our results suggest that twin structure modulates $T_{c}$ by modulating the local dielectric environment. Discriminating between theories of superconductivity in STO will require a microscopic understanding of the consequences of structurally driven local variations in the dielectric properties of STO. 

\section{\label{sec:conclusion}Conclusion}
We have shown that tetragonal domain structure locally enhances the superconducting transition temperature in regions of two-dimensionally doped STO. While it is not surprising that changes in the crystal lattice affect $T_{c}$, our observation that $T_{c}$ is enhanced by $\agt$ 10\% while the lattice constants change by only 0.1\% is notable, and suggests that the dielectric properties of STO play an important role in this material's superconductivity. 

The modulation in $T_{c}$ that we detected in two-dimensionally doped STO is likely relevant in systems in which superconductivity arises due to interface effects between STO and another material, such as LAO/STO \cite{OhtomoNat04,ReyrenSci07} and monolayer FeSe grown on STO \cite{HeNatMater13,TanNatMater13,LeeNat14,GeNatMater14}. Our results further motivate the development of microscopic modeling of STO that takes structure as well as local dielectric properties into account.

\begin{acknowledgments}
We thank John Berlinsky, Robert Laughlin, Steven Kivelson, Srinivas Raghu, and Akash Maharaj for helpful discussions and Christopher Watson for feedback on our manuscript. This work was supported by the Department of Energy, Office of Science, Basic Energy Sciences, Materials Sciences and Engineering Division, under Contract DE-AC02-76SF00515. Differential aperture X-ray microdiffraction was carried out at the Advanced Photon Source, a DOE Office of Science User Facility operated for the DOE Office of Science by Argonne National Laboratory under Contract No. DE-AC02-06CH11357. H. N. acknowledges support from a Stanford Graduate Fellowship and a Natural Sciences and Engineering Council of Canada PGS D. T.A.M. also acknowledges support from the National Science Foundation Graduate Research Fellowship under Grant No. DGE-114747.
\end{acknowledgments}

\appendix

\section{\label{sec:ns}Connection between susceptibility and superfluid density}
We report susceptibility in units of superconducting flux quanta per ampere of current passing through the field coil, ${\Phi}_{0}$/A, where ${\Phi}_{0}$ = $h$/(2$e$),  $h$ is Planck's constant, and $e$ is the electron charge.

Under certain conditions, the susceptibility signal measured with our scanning SQUID is directly proportional to the superfluid density, $n_{s}$. The superconductor must be in the Pearl limit, with the superconducting thickness, $d_{sc}$, much smaller than the penetration depth, $\lambda$ \cite{PearlAPL64}. Additionally, the field coil diameter and the distance between the SQUID and the superconductor must be much larger than $d_{sc}$ \cite{KirtleyPRB12}. If these conditions are satisfied, then the susceptibility signal at constant height is inversely proportional to the Pearl length, $\Lambda$= $2{\lambda}^{2}/d_{sc}$, and is proportional to the superfluid density $n_{s}$ = $2m^{\star}/{\mu}_{0}e^{2}{\Lambda}$, where $m^{\star}$ is the effective mass \cite{KirtleyPRB12,BertPRB12}. 

Our measurements at temperatures below the emergence of separated domains of diamagnetism were in this limit: the thickness of the Nb-doped layer in the $\delta$-doped STO was 5.5 nm for the data discussed in the main text (Fig.~\ref{fig:1}, Fig.~\ref{fig:2}, Fig.~\ref{fig:3}, Fig.~\ref{fig:4}) (36.9 nm in Appendix~\ref{sec:similar features}). The superconducting thickness in the $\delta$-doped STO, estimated from the temperature dependence of the upper critical field, was somewhat larger than the thickness of the doped layer (for example, in the $d$ = 5.5 nm sample, the estimated thickness of the superconductivity was 8.4 nm \cite{KozukaNat09}) but still within the same limit. The Pearl length estimated from fits to Pearl vortices (not shown) was on the order of 600-800 $\mu$m, much larger than the thickness of the superconductivity. The field coil diameter was $\approx$ 20 $\mu$m, and the distance from the SQUID to the sample was 1-2 $\mu$m. 

\section{\label{sec:microdiff}X-ray microdiffraction}

\begin{figure}
\includegraphics{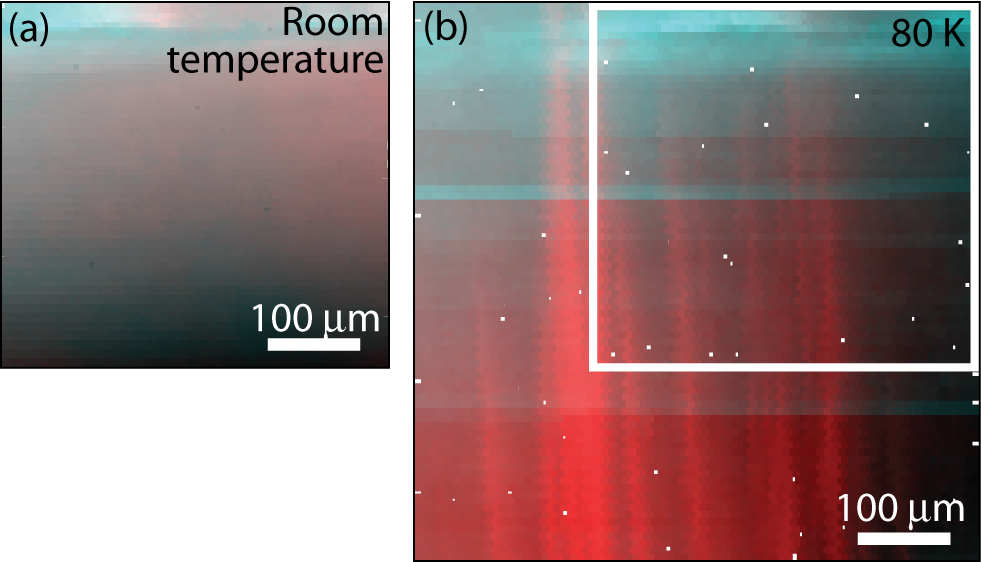}
\caption{\label{fig:microdiff}Color. Tilt map of the 1 at.\% Nb $\delta$-doped STO reveals domain structure below 105 K. Changes in color indicate changes in the local orientation of the crystal lattice relative to the incident X-ray beam. Red, green, and blue correspond to the $x$, $y$, and $z$ components, respectively, of displacements of the $c$* reciprocal lattice vector relative to a reference position. The intensity of a particular color reflects the magnitude of the displacement relative to the maximum displacement in that channel in the entire scan area. The scan plane was parallel to the surface of the sample. (a) At room temperature, the lattice is relatively uniform. (b) In contrast, at 80 K, below the cubic-to-tetragonal structural transition, the sample displays features whose orientation and size are consistent with tetragonal domain structure. The white box in (b) indicates the position of the room temperature image in (a). The horizontal feature at the top of (a) and (b) did not change with thermal cycling and is likely due to physical damage to the sample.}
\end{figure}

We performed differential aperture X-ray microdiffraction \cite{microdiff1,microdiff2,microdiff3} experiments at beamline 34-ID-E at the Advanced Photon Source, Argonne National Laboratory \cite{Xraypaper} on the 1 at.\% Nb $\delta$-doped STO. This beamline is equipped with a liquid nitrogen-cooled stage that we used to cool the STO below its structural transition temperature of 105 K \cite{UnokiJPSJ67}. We collected Laue diffraction patterns while rastering the sample under the X-ray beam (beam width $\approx$ 1 $\mu$m, planar step size $\approx$ 5 $\mu$m), then indexed each pattern to a distorted room temperature cubic unit cell for STO \cite{microdiff1,microdiff2,microdiff3} in order to determine the orientation of the local crystal structure. 

After collecting Laue diffraction patterns in a rastered grid of points, we indexed each pattern to the room temperature orientation of the cubic unit cell for STO. To create a spatial map of tilts of the unit cell, we expressed the orientation of the unit cell in terms of vectors in a three-dimensional, rectangular space. Since the primitive lattice vectors are orthogonal to one another in a cubic or tetragonal unit cell, the orientation of any one of these vectors relative to a fixed coordinate system uniquely describes the orientation of the entire unit cell, thus encoding tilting of the unit cell.

In the images in Fig.~\ref{fig:microdiff}, we represent changes in the orientation of the lattice by encoding the ($x$,$y$,$z$) components of the $c$* reciprocal lattice vector in red, green, and blue, respectively. At room temperature, above the cubic-to-tetragonal structural phase transition temperature, the lattice orientation changes smoothly and by very little over hundreds of microns [Fig.~\ref{fig:microdiff}(a)]. In contrast, at 80 K, below the cubic-to-tetragonal transition, there are abrupt changes in tilt whose orientations and sizes are consistent with tetragonal domains [Fig.~\ref{fig:microdiff}(b)].

\section{\label{sec:similar features}Similar features in another {\lowercase{$\delta$}}-doped sample}

\begin{figure}
\includegraphics{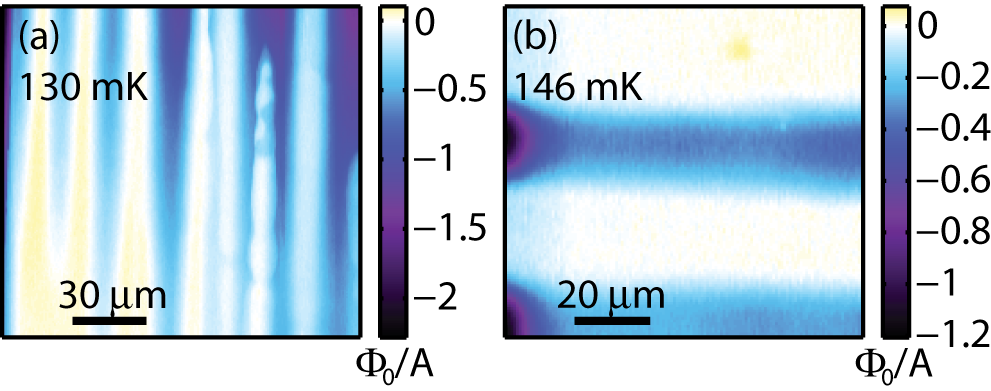}
\caption{\label{fig:similar features}Color. Regions of enhanced $T_{c}$ in 0.2 at.\% Nb $\delta$-doped STO. The scans shown in (a) and (b) were taken in different areas of the same sample.}
\end{figure}

We measured a second $\delta$-doped sample with $d$ = 36.9 nm , $N_{D}$ = 0.2 at.\% Nb. We detected long, narrow regions of diamagnetism surrounded by paramagnetism (Fig.~\ref{fig:similar features}), similar to our observations in the $d$ = 5.5 nm, $N_{D}$ =  1 at.\% Nb sample (Fig.~\ref{fig:1} of the main text). The temperature scale for superconductivity in the 0.2 at.\% sample was lower than for the 1 at.\% sample, consistent with global resistance measurements (inset to Fig.~\ref{fig:Tcs}) made in a separate cooldown on the two samples.

\section{\label{sec:widths}Widths of the stripes}

\begin{figure*}
\includegraphics{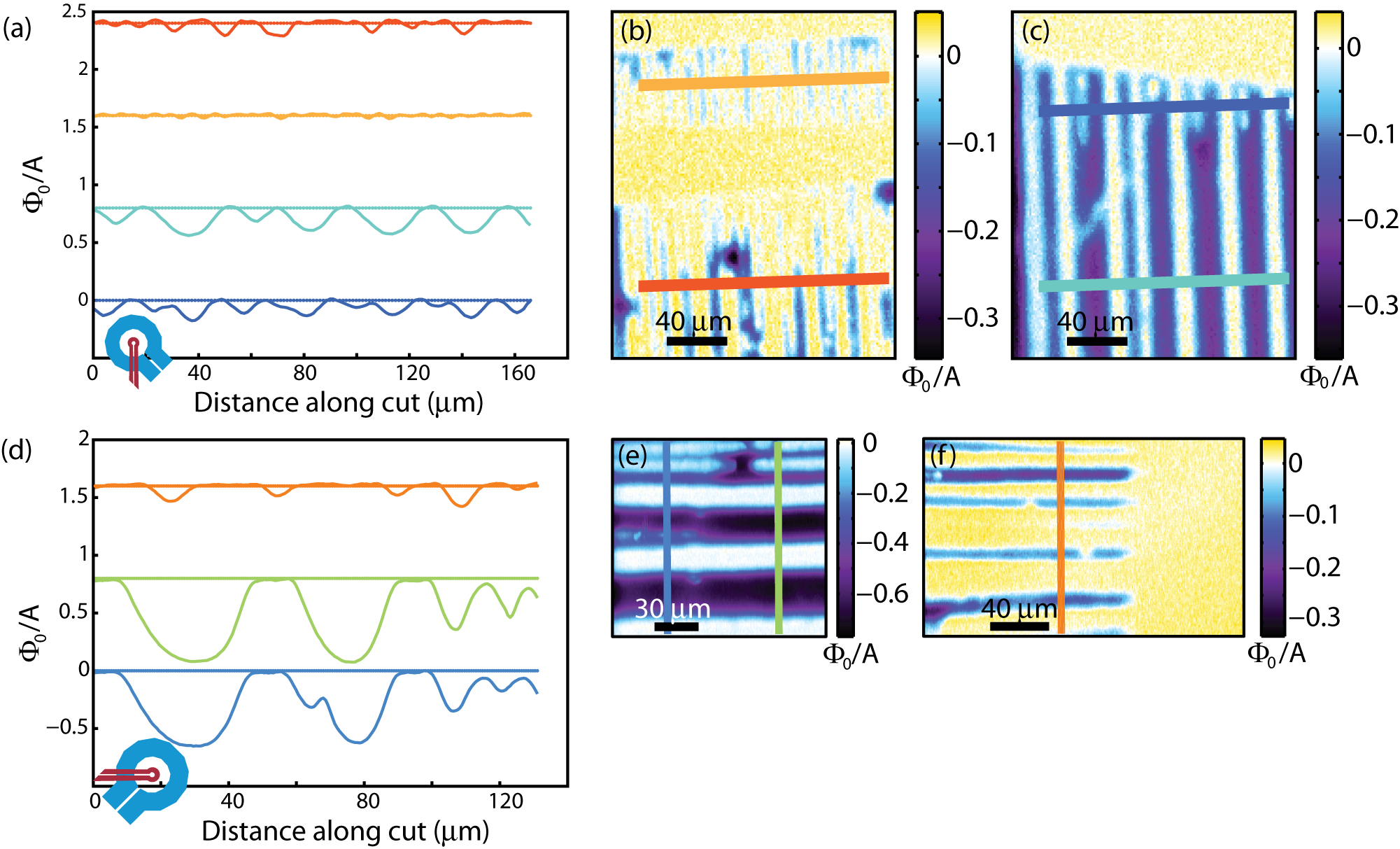}
\caption{\label{fig:widths}Color. We detected considerable variation in the widths of the regions of enhanced $T_{c}$, indicating that at least some of the features are not resolution-limited. (a) Averaged linecuts of the vertical features in (b) and (c). (d) Averaged linecuts of the horizontal features in (e) and (f). The curves in (a) and (d) are offset by intervals of 0.8 ${\Phi}_{0}$/A for clarity. Dotted lines indicate zero susceptibility for each curve. The schematics of the SQUID pickup loop (red) and field coil (blue) in (a) and (d) are to scale and are oriented relative to the linecuts as they were during data acquisition. All images were taken at 320 mK and are of the $\delta$-doped sample with 1 at.\% Nb doping.}
\end{figure*}

An upper bound on the spatial resolution of our susceptibility measurements would be set by the length scale of variations in the field that we apply to the sample (by the diameter of the field coil, which was $\approx$ 20 $\mu$m in the present investigation). This scenario would be relevant, for example, in the case of a three-dimensional superconductor with a penetration depth much smaller than our sensor size and with a correspondingly strong diamagnetic response. The system that we studied was in a very different limit; we studied a two-dimensional superconductor near $T_{c}$ in which the diamagnetic response was weak, with the response field produced by the sample being a factor of $10^{3}$ smaller than the maximum applied field. In this limit, we expect the spatial resolution of our susceptibility measurements to be smaller than the length scale of the field coil. A lower bound on the resolution of our susceptibility images is set by the diameter of the pickup loop ($\approx$ 3 $\mu$m) and the distance between our sensor and the sample \cite{NowackNatMater13}. 

If the underlying source of the features that we observed were narrower than our spatial resolution, e.g. domain boundaries or very narrow domains, then we would expect to see many features of the same apparent width in our images. On the other hand, if the underlying source were sometimes wider than our spatial resolution, then we would expect to see a range of widths in our susceptibility images. We observe a variety of widths, many of which were wider than our pickup loop, and a few of which were even wider than the diameter of our field coil (Fig.~\ref{fig:widths}).

\section{\label{sec:Tcs}Lower and upper {$T_{c}$}}

\begin{table*}
\caption{\label{tab:Tcs}Estimates of lower and upper $T_{c}$ and ${\Delta}T_{c}/T_{c}$ for regions of the 1 at.\% $\delta$-doped sample displayed in the main text.}
\begin{ruledtabular}
\begin{tabular}{lllllp{5cm}}
Fig. & $T_{c,\textnormal{low}} (mK)$ & $T_{c,\textnormal{high}}$ (mK) & ${\Delta}T_{c}/T_{c}$ (\%) & Scanned area (${\mu}m^{2}$)& Notes\\
\hline
~\ref{fig:1}(a) & 310 & 350 & 13 & $4.2{\times}10^{4}$& \\
~\ref{fig:1}(b) & 300 & 340 & 13 & $4.2{\times}10^{4}$ & Small region remained diamagnetic up to at least 370 mK\\
~\ref{fig:1}(c) & 300 & 340 & 13 & $2.9{\times}10^{4}$ & \\
~\ref{fig:2}(b) & 320 & 380 & 19 & $4.4{\times}10^{4}$ & Scan area overlapped with region that was masked with clip during growth; small regions remained diamagnetic up to at least 400 mK\\
~\ref{fig:3}(a) & 320 & 370 & 16 & $1.9{\times}10^{4}$ & \\
~\ref{fig:3}(b) & 320 & 350 & 9 & $1.9{\times}10^{4}$ & Small region remained diamagnetic up to at least 380 mK\\
~\ref{fig:4}(a) & 320 & 360 & 13 & $1.9{\times}10^{4}$ & Small region remained diamagnetic up to at least 380 mK\\
\end{tabular}
\end{ruledtabular}
\end{table*} 

\begin{figure}
\includegraphics{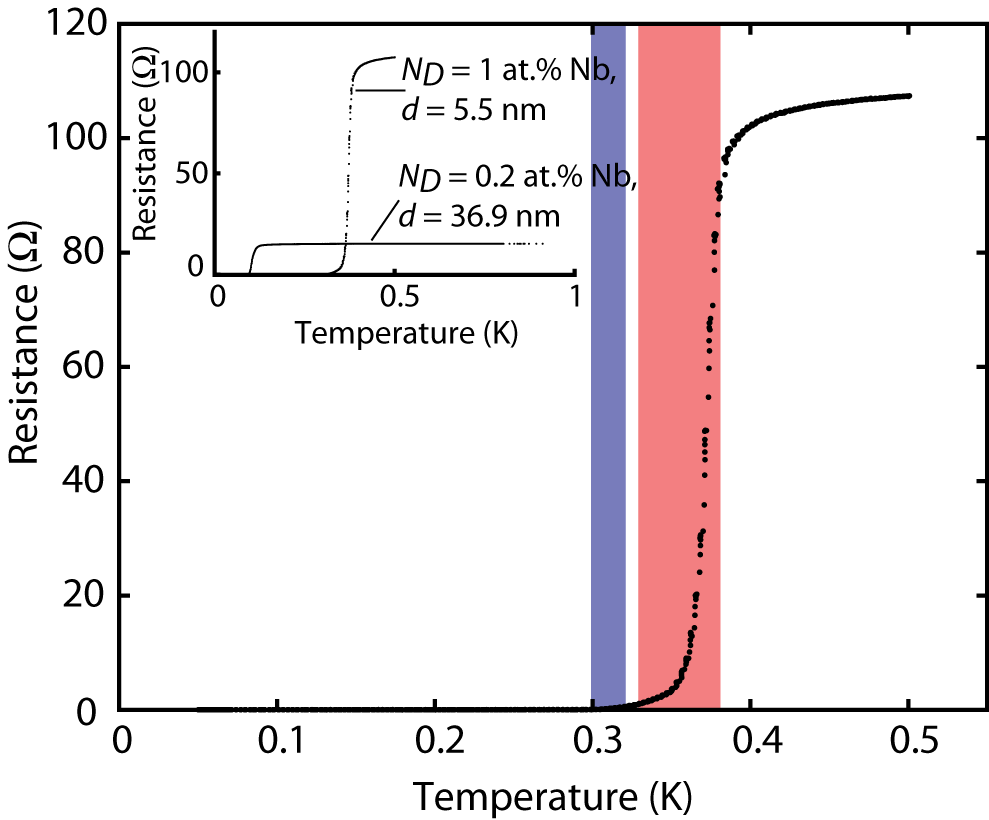}
\caption{\label{fig:Tcs}Color. Comparison of superconducting transitions measured in global transport and local susceptibility for the $\delta$-doped sample with $N_{D}$ = 1 at.\% Nb and $d$ = 5.5 nm shows that zero resistance occurs just below the lower $T_{c}$ determined from local susceptibility. Resistance as a function of temperature measured in a separate cooldown is shown in black dots; ranges of lower and upper transition temperatures inferred from maps of susceptibility (and reported in Table~\ref{tab:Tcs}) are shown by blue and red bands, respectively. Inset: Resistance as a function of temperature for the two $\delta$-doped samples studied shows that the overall temperature scale for superconductivity in the sample with $N_{D}$ = 0.2 at.\%, $d$ = 36.9 nm is lower than for the $N_{D}$ = 1 at.\% Nb, $d$ = 5.5 nm sample.}
\end{figure}

We calculated the percent variation in $T_{c}$ according to ${\Delta}T_{c}/T_{c}$ = 100\%$\times$($T_{c,\textnormal{high}}-T_{c,\textnormal{low}}$)/$T_{c,\textnormal{low}}$. For the purposes of these estimates, we defined $T_{c}$ conservatively. For example, if a scan at 370 mK still showed regions of diamagnetism but 380 mK did not, we took $T_{c,\textnormal{high}}$ = 370 mK. For $T_{c,\textnormal{low}}$, if a featureless area contained some patchy normal regions at 310 mK but was not fully normal until 320 mK, we assigned $T_{c,\textnormal{low}}$ = 320 mK [Fig.~\ref{fig:3}(a)]. In Table~\ref{tab:Tcs}, we summarize values of $T_{c,\textnormal{low}}$, $T_{c,\textnormal{high}}$, and ${\Delta}T_{c}/T_{c}$ for the regions displayed in Fig.~\ref{fig:1}(a-c), Fig.~\ref{fig:2}(b), Fig.~\ref{fig:3}(a-b), and Fig.~\ref{fig:4}(a) of the main text. In Fig.~\ref{fig:Tcs}, we compare the transition temperatures determined from susceptibility scans that are tabulated in Table~\ref{tab:Tcs} to a global measurement of resistance vs. temperature made on the same sample in a separate cooldown. 

With the exception of the images presented in Fig.~\ref{fig:similar features}, the temperatures reported for the scanning SQUID measurements were measured at the mixing chamber of our dilution fridge. In an earlier cooldown, we recorded the temperature at our scanner cage using a ruthenium oxide thermometer in a copper bobbin that was rigidly mounted to the oxygen-free, high-conductivity copper cage. Above 100 mK, we found that the mixing chamber temperature was a reasonable proxy for the cage temperature to approximately $\pm$10 mK. Temperatures reported in Fig.~\ref{fig:similar features} were measured with the ruthenium oxide thermometer.

\section{\label{sec:location}Location of enhancement of {$T_{c}$}}

\begin{figure}
\includegraphics{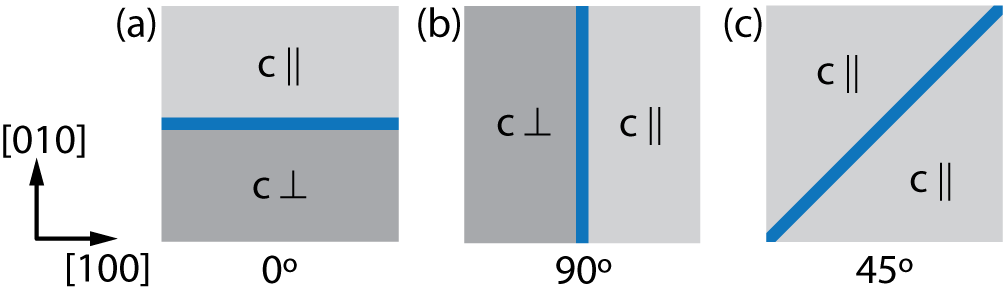}
\caption{\label{fig:location1}Allowed orientations of boundaries between tetragonal domains as seen in a cubic (001) plane. Arrows indicate the former cubic [100] and [010] directions. Boundaries between a domain with its tetragonal $c$ axis parallel to the plane and a domain with $c$ axis perpendicular to the plane are at (a) 0$^{\circ}$ or (b) 90$^{\circ}$ to the cubic [100] direction. (c) Boundaries between two domains with $c$ axis parallel to the plane are at 45$^{\circ}$ to the cubic [100] direction.}
\end{figure}

\begin{figure*}
\includegraphics{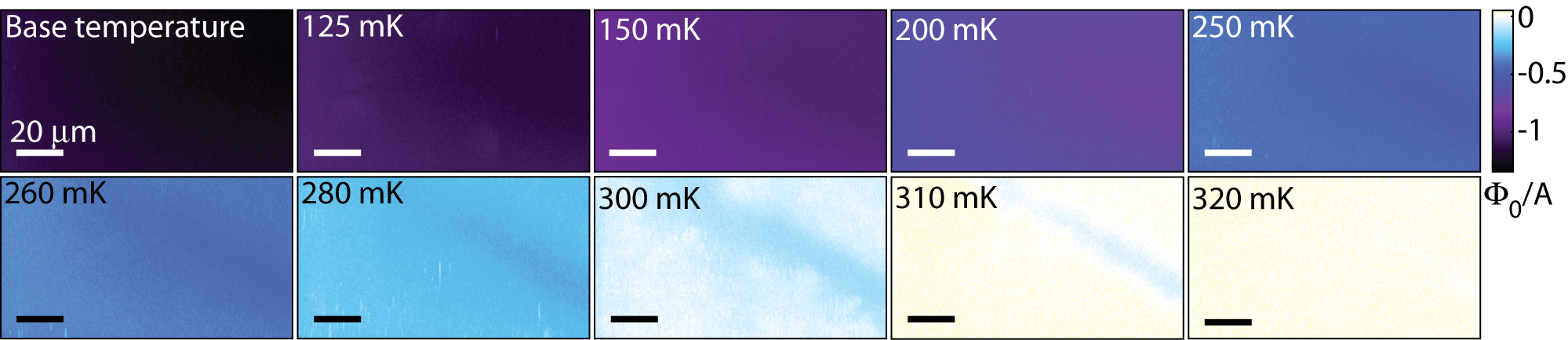}
\caption{\label{fig:location2}Color. A single feature oriented at -45$^{\circ}$ to the former cubic [100] axis exhibits higher $T_{c}$ than its surroundings. Repeated scans at a single location on the 1 at.\% Nb $\delta$-doped sample at the temperatures indicated on the images.}
\end{figure*}

In the first cooldown, we imaged the temperature dependence of the susceptibility in ten regions, corresponding to approximately 4\% of the total sample area, and found only one feature oriented at (-)45$^{\circ}$ to cubic [100] [probably corresponding to the case illustrated in Fig.~\ref{fig:location1}(c)] whose $T_{c}$ was clearly enhanced relative to its surroundings (Fig.~\ref{fig:location2}). In the second cooldown, we imaged the temperature dependence of the susceptibility in six regions, corresponding to approximately 2\% of the total sample area, and did not find any features of enhanced $T_{c}$ at (-)45$^{\circ}$ to cubic [100]. 

\section{\label{sec:3d}Bulk doped STO}

\begin{figure}
\includegraphics{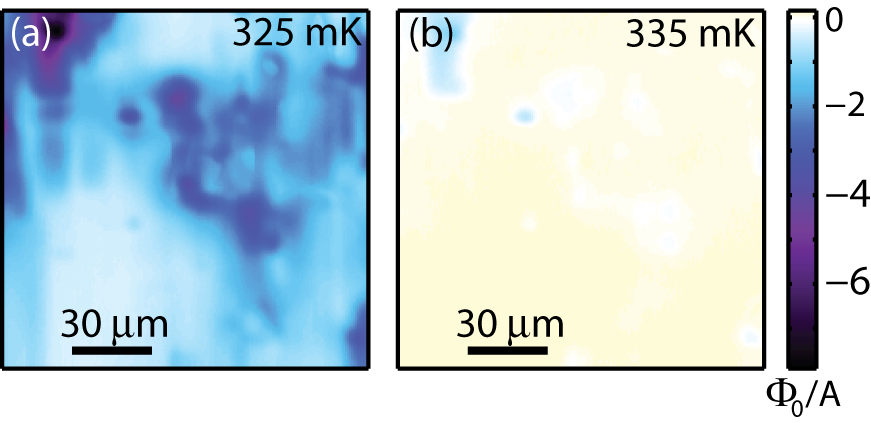}
\caption{\label{fig:3d}Color. Susceptibility images of bulk 1 at.\% Nb-doped STO near $T_{c}$ reveal faint regions of enhanced diamagnetic response oriented along cubic axes. (a,b) Scans in the same area at (a) 325 mK and (b) 335 mK. At 325 mK, the entire image area still is superconducting. The band of faint features in the upper half of (a) is oriented along cubic [100] but does not persist above the “bulk” $T_{c}$ of its surroundings.}
\end{figure}

To check whether single-crystal, bulk-doped STO exhibited similar features of enhanced $T_{c}$, we mapped susceptibility as a function of temperature in a single-crystal sample of 1 at.\% Nb-doped STO. We observed faint rectangular features of stronger diamagnetic response surrounded by comparatively weaker diamagnetism near, but fully below, $T_{c}$ [Fig.~\ref{fig:3d}(a)]. These features did not persist above the $T_{c}$ of their surroundings (to within our temperature step size of 10 mK) [Fig.~\ref{fig:3d}(b)].

\bibliography{references}

\end{document}